\def\simless{\mathbin{\lower 3pt\hbox
     {$\rlap{\raise 5pt\hbox{$\char'074$}}\mathchar"7218$}}}   
\def\simmore{\mathbin{\lower 3pt\hbox
     {$\rlap{\raise 5pt\hbox{$\char'076$}}\mathchar"7218$}}}   
\def\Msun{{\rm M}_\odot}                                       

\def\EXO{EXO 0748$-$676 }

\documentclass[useAMS,usenatbib]{mn2e}

\usepackage{threeparttable} 
\usepackage{subfigure}
\usepackage{graphicx}

\title[EXO 0748--676]{The distance and internal composition of the neutron star in 
\EXO with XMM-Newton}

\label{firstpage}

\date{Accepted. Received; in original form}

\author[G. Zhang et al.]{Guobao Zhang$^{1}$\thanks{E-mail:
zhang@astro.rug.nl}, Mariano M\'endez$^{1}$ , Peter Jonker$^{2,3,4}$ and Beike Hiemstra$^{1}$.\\
$^{1}$Kapteyn Astronomical Institute, University of Groningen, P.O. BOX
800, 9700 AV Groningen, The Netherlands\\
$^{2}$SRON, Netherlands Institute for Space Research, Sorbonnelaan 2, 3584~CA, Utrecht, The Netherlands\\
$^3$Harvard--Smithsonian Center for Astrophysics, 60 Garden Street, Cambridge, MA~02138, U.S.A.\\
$^4$Department of Astrophysics, IMAPP, Radboud University Nijmegen, PO Box 9010, NL-6500 GL Nijmegen, the Netherlands\\}

\begin{document}


\maketitle

\label{firstpage}

\date{Accepted. Received; in original form}

\begin{abstract}

Recently, the neutron star X-ray binary EXO 0748-676 underwent a transition to quiescence. We analyzed an XMM-Newton observation of this source in quiescence, where we fitted the spectrum with two different  neutron-star atmosphere models. From the fits we constrained the allowed parameter space in the mass-radius diagram for this source for an assumed range of distances to the system. Comparing the results with different neutron-star equations of state, we constrained the distance to EXO 0748-676. We found that the EOS model 'SQM1' is rejected by the atmosphere model fits for the known distance, and the 'AP3' and 'MS1' is fully consistent with the known distance.   

\end{abstract}


\section{Introduction}

The low-mass X-ray binary (LMXB) EXO 0748--676 was discovered as a transient 
source with the European X-ray Observatory Satellite ({\em EXOSAT}) in 1985 
\citep{Parmar86}. The source exhibits simultaneous 
X-ray and optical eclipses from which the orbital period of P = 3.82 hr was deduced
\citep{Crampton86}. \EXO also exhibited irregular X-ray dipping activity \citep{Parmar86},
and type-I X-ray bursts \citep{Gottwald86}. Burst oscillations 
in \EXO were first reported by \cite{Villarreal04} at 45 Hz in the average Fourier Power Spectrum of 38 
type-I X-ray bursts; the 45-Hz signal was then interpreted as the spin frequency of 
the neutron star. Recently, \cite{Galloway09} detected  millisecond oscillations in 
the rising phase of two type-I X-ray bursts in EXO 0748-676 at a frequency of 552 Hz. 
They concluded that the spin frequency of EXO 0748-676 is close to 522 Hz,
rather than 45 Hz as suggested by \cite{Villarreal04}.  The 45 Hz oscillation 
may arise in the boundary layer between the disk and the neutron star \citep{Balman09} 
or it could be a statistical fluctuation \citep{Galloway09}. \cite{Cottam02} reported a 
measurement of the gravitational redshift from iron and oxygen  X-ray absorption lines arising from the atmosphere of the neutron star in \EXO during type-I X-ray bursts, but
subsequent observations failed to confirm these features \citep{Cottam08}. Based on the 
gravitational redshift, \cite{Ozel06} suggested that the mass, radius  and distance 
of \EXO are $2.10 \pm 0.28$ $\Msun$, $13.8 \pm 1.8$ km and $9.2 \pm 1.0$ kpc, respectively, 
which would rule out many neutron-star equations of state.

Measuring the distance to LMXBs is difficult,
except for the case of sources in globular clusters. A way to get the distance is 
using type-I X-ray bursts. The peak flux for some very bright bursts can reach the 
Eddington luminosity at the surface of the neutron star. From a strong X-ray burst,
\cite{Wolff05} derived a distance to \EXO of 7.7 kpc for a helium-dominated burst 
photosphere, and  5.9 kpc for a hydrogen-dominated burst photosphere. 
\cite{Galloway08a} analyzed several type-I X-ray bursts from \EXO and estimated a 
distance of  7.4 kpc, different from the value of 9.2 kpc reported by \cite{Ozel06}.
 Taking into account the touchdown flux and high-inclination in 
\EXO, recently \cite{Galloway08b} gave a distance of 7.1 $\pm$ 1.2 kpc.

Another way to get the distance to an LMXB is through observations of quiescent X-ray 
emission from the neutron-star surface. During the quiescent state, X-ray emission 
originates from the atmosphere of the neutron star. By fitting the  X-ray 
spectrum of the neutron-star system with hydrogen atmosphere models, one can 
estimate the mass, radius and distance of the neutron star.  Recently, the neutron-star
 X-ray transient \EXO underwent a  transition into quiescence 
\citep{Degenaar09,Bassa09}. 

In this paper, we report on the distance to \EXO that we constrained from XMM-Newton data. 
We use two different neutron-star atmophere models to fit the X-ray spectrum, and compare the 
results of the spectral fitting with different neutron-star equation of state (EOS).
In the next section, we describe the observation and data analysis. We show the fitting 
results in \S\ref{results},  and  we discuss our findings in \S\ref{discussion}.

\section[]{OBSERVATIONS AND DATA ANALYSIS}
\label{data}
\EXO was observed with the European Photon Imaging Camera (EPIC PN and MOS) on board the 
XMM-Newton on 2008 November 6 at 08:30:03 UTC (obsID  0560180701). The PN and the two MOS 
cameras were operated in Full-Window mode. We reduced the XMM-Newton Observation 
Data Files (ODF) using version 8.0.0 of the science analysis software (SAS). We used the {\sc epproc} and 
{\sc emproc} tasks to extract the event files for the PN and the two MOS 
cameras, respectively. Source light curves and spectra were extracted in the 0.2 $-$ 12.0 keV band  
using a circular extraction region with a radius of 30 arcsec centered on the 
position of the source.
Background light curves and spectra were  extracted from a circular source-free region 
 of 35 arcsec source-free on the same CCD. We applied standard filtering and examined the 
light curves for background flares. No flares were present and we used the 
whole exposure for our analysis. 
The exposure time for the PN camera was 24.2 ks, and for each MOS camera was 29.03 ks.  
The source count rate was $0.496\pm 0.005$ cts/s for PN,  and  
 $0.135\pm 0.002$ cts/s and $0.127\pm 0.002$ cts/s for MOS1 and MOS2, respectively.
We checked the filtered event files for 
photon pile-up by running the task {\sc epatplot}. No pile-up was apparent 
in the PN, MOS1 and MOS2 data. The photon redistribution matrices and 
ancillary files for the source spectra were created using the SAS tools 
{\sc rmfgen} and {\sc arfgen}, respectively. We rebinned the source spectra 
using the tool {\sc pharbn}\footnote{M. Guainazzi, private communication}, 
such that the number of bins per resolution element of the PN and MOS spectra 
was 3 and the minimum number of counts per channel was 20.

We fitted the PN and MOS spectra simultaneously in the 0.5$-$10.0 keV range with {\sc XSPEC} 12.50 \citep{Arnaud96}, using either of 
two neutron-star hydrogen-atmosphere models: NSAGRAV \citep{Zavlin96} and
NSATMOS \citep{Heinke06}. The NSAGRAV model provides the spectra emitted from a nonmagnetic
hydrogen atmosphere of a neutron star with surface gravitational acceleration, $g$, ranging
from $10^{13}$ to $10^{15}$ cm s$^{-2}$. This model uses the mass ($M_{\rm NS}$) and radius 
($R_{\rm NS}$) of the neutron
star and the unredshifted effective temperature of the surface of the star ($kT_{\rm eff}$)
as parameters. The normalization of the model is defined as $1/D^2$, where $D$ is the
distance to the source in pc. The second model that we used, NSATMOS, includes 
a range of surface gravities and effective temperatures, and incorporates thermal 
electron conduction and self-irradiation by photons from the compact object.
This model assumes negligible magnetic fields (less than $10^9$ G) 
and a pure hydrogen atmosphere. NSATMOS parameters are $M_{\rm NS}$, $R_{\rm NS}$, log$T_{\rm eff}$ (the same as for NSAGRAV), distance in kpc, and a separate normalization $K$, which corresponds to the fraction of the neutron-star surface that is emitting. We fixed $K$ to be 1 in
all our fits with NSATMOS.

\begin{figure}
\centering
\includegraphics[width=65mm,angle=270]{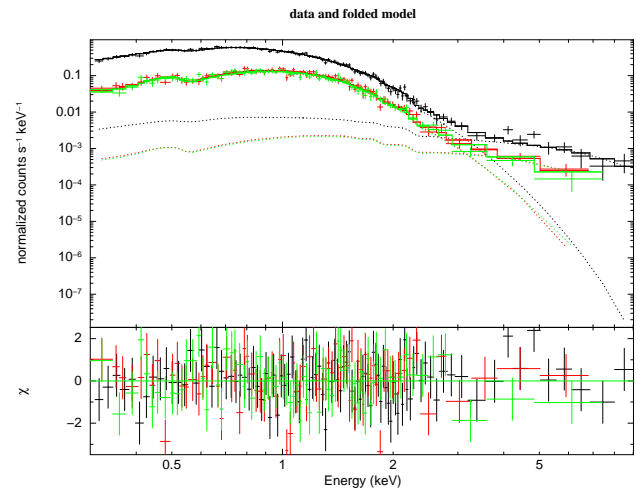}
\caption{XMM-Newton PN (black), MOS1 (red) and MOS2 (green) spectrum of \EXO in the 0.5$-$
10.0 keV energy band. The spectrum was fitted with a neutron-star hydrogen atmosphere model (NSATMOS) and a power-law model with $\Gamma$ fixed to 1.  The lower panel shows the residuals to the best-fit model. } 
\label{spectrum}
\end{figure}

We included the effect of interstellar absorption using PHABS assuming cross-sections of 
\cite{Balucinska92} and solar abundances from \cite{Anders89}, and we
let $N_{\rm H}$, column density along the line of sight free to vary during the fitting.
In order to account for  differences in  effective area between  the different cameras,
we introduced a multiplicative factor in our model. First, this factor was fixed to unity for PN and 
free for MOS1 and MOS2.  Then, we set the scaling factor to unity for MOS1 and MOS2, respectively, 
and set the factor free for the other cameras.  We found that, fixing the
scaling factor for different cameras gives similar best-fit results. Therefore in the rest of the 
paper we fixed the factor to be 1 for PN and free to vary for the other cameras.
None of the atmosphere models alone fitted the spectrum above $\sim 2-3$ keV 
properly. Adding a power-law component improved the fits significantly, however, 
all parameters were less constrained than when fitting the data with the neutron-star 
atmosphere model only. 
We first fixed the power-law index to 0.5, 1.0 and 1.5 to get better constraints on the  parameters  of the neutron-star atmosphere model (Degenaar et al. 2009). Further, we initially fixed the distance to the NS at 7.1 kpc, which is the value inferred from the touchdown flux of \cite{Galloway08b} .

\section[]{RESULTS}
\label{results}

\subsection[]{Results from the spectral fits}
\label{fitting_results}

\begin{table*}
\begin{center}
\caption{Best-fit parameters of neutron-star atmosphere models fit to the XMM-Newton data of \EXO. }

\begin{tabular}{ccccccccccc}

\hline \hline
model   & N$_{\rm H}$   & $T^{\infty}_{\rm eff}$    & $M_{\rm NS}$   & $R_{\rm NS}$ & 
$\Gamma$ & $F_{pow}$ &$F_X$ &$\chi^{2}$/d.o.f. \\
        &($10^{20} cm^{-2}$)&(eV)  &($\Msun$)  &(km) & & 10$^{-13}$ ergs cm$^{-2}$ s$^{-1}$& 10$^{-12}$ ergs cm$^{-2}$ s$^{-1}$  &  \\
\hline
NSAGRAV     &$5.6\pm 1.8$ & $113^{+14}_{-8}$ & $1.55\pm 0.18$  & $15.2\pm 1.8$  & 0.5 &$1.15\pm 0.21$   & $1.18\pm 0.15$   & 0.986/219  \\
\hline 
NSATMOS     &$5.4\pm 1.5$ & $113\pm 4$ & $1.29\pm 0.20$ & $16.1^{+0.9}_{-1.2}$  & 0.5 & $1.17\pm 0.20$ & $1.23\pm 0.16$ & 0.985/219  \\

\hline
NSAGRAV     &$6.2^{+1.3}_{-1.8}$ & $114^{+24}_{-3}$ & $1.62\pm 0.11$  & $15.8^{+0.25}_{-3.5}$  & 1.0 &$1.10\pm 0.15$   & $1.14\pm 0.13$   & 0.977/219  \\
\hline 
NSATMOS     &$6.1\pm 1.5$ & $114\pm 4$ & $1.55\pm 0.12$ & $16.0^{+0.7}_{-1.3}$  & 1.0 & $1.11\pm 0.15$ & $1.13\pm 0.06$ & 0.977/219  \\

\hline
NSAGRAV     &$6.7\pm 1.5$ & $110\pm 8$ & $1.71\pm 0.30$  & $16.5\pm 0.5$  & 1.5 &$1.00\pm 0.19$   & $1.01\pm 0.15$   & 0.987/219  \\
\hline 
NSATMOS     &$6.7\pm 1.4$ & $110\pm 5$ & $1.77\pm 0.45$ & $16.6^{+1.8}_{-7.5}$  & 1.5 & $1.03\pm 0.22$ & $1.03\pm 0.10$ & 0.985/219  \\

\hline 
\label{tab:model}
\end{tabular}
\begin{tablenotes}
\item[]Note. --  N$_{\rm H}$  is the equivalent hydrogen column density, $T^{\infty}_{\rm eff}$ the effective temperature of the neutron-star surface as seen at infinity,  $M_{\rm NS}$ and $R_{\rm NS}$ are the mass and radius of the neutron star, respectively.  $F_{pow}$ is the unabsorbed flux of the power-law component in the 0.5$-$10 keV energy band, and $F_X$ is the total unabsorbed X-ray flux in the same energy band. The last column gives the reduced $\chi^{2}$ for 219 degrees of freedom.
The quoted errors represent the 90\% confidence levels. 
\end{tablenotes}
\end{center}
\end{table*}

Figure \ref{spectrum} shows the XMM-Newton spectra of EXO 0748-676 fitted with the model ``phabs (NSATMOS + powerlaw) ''.  The power-law index is fixed at 1.0.  
The best fit of this model gives 
$N_{\rm H} = 6.1\pm 1.5$ $\times 10^{20}$ cm$^{-2}$, neutron-star mass
 $M_{\rm NS} = 1.55 \pm 0.12 \Msun$, neutron-star radius 
$R_{\rm NS} = 16.0^{+0.7}_{-1.3}$ km, and effective temperature log$T_{\rm eff} = 6.20\pm 0.02$ (in $K$). 
According to the same formula $T^{\infty}_{\rm eff}= T_{\rm eff}\sqrt{1-(2GM_{\rm NS})/(R_{\rm NS}c^2)}$ used by \cite{Degenaar09},
we converted $T_{\rm eff}$ to the effective temperature as seen by an observer at infinity,  $T^{\infty}_{\rm eff}=114\pm 4$ eV.  
In the formula, $G$ is the gravitational constant and $c$ is the speed of light.
The model predicts 0.5$-$10 keV an unabsorbed X-ray flux   
$F_X = 1.13\pm 0.06 \times 10^{-12}$ ergs cm$^{-2}$ s$^{-1}$. The  
flux of the power-law component in the same energy band is   
$F_{pow} = 1.11\pm 0.15 \times 10^{-13}$ ergs cm$^{-2}$ s$^{-1}$,
which corresponds to $\sim 10 \%$ of the total unabsorbed flux.  
The reduced $\chi^2$ is 0.977 for 219 degrees of freedom. The best-fit results of the models 
NSAGRAV and NSATMOS for the three different power-law 
index are given in Table-\ref{tab:model}. Errors
are given at the 90\% confidence level for  one fit parameter.

We  note from Table \ref{tab:model} that both atmosphere models, regardless of the value of $\Gamma$, yield a good fit with similar $\chi^2$. In the rest of the analysis, we used a power-law index fixed to 1. Further, $N_{\rm H}$ and $T_{\rm eff}$ are well constrained and are consistent for the different fits. 
Both NSAGRAV and NSATMOS models also give consistent results on  $M_{\rm NS}$ and $R_{\rm NS}$.
The  NSATMOS model
is more accurate in constraining $T_{\rm eff}$ than the NSAGRAV model.

\subsection[]{Equation of state}
\label{EOS_results}

Fitting the quiescence XMM-Newton spectrum of \EXO  with two different atmosphere models and comparing the results allows us to test the reliability and accuracy of both models. From the fits we get a mass and radius of the neutron star at a specified distance, and then by comparing the inferred mass and radius with the different neutron-star EOS we can give upper limits to the source distance for the different EOS.

\begin{figure*}
\begin{center}
\centering
    \subfigure[5.0 kpc]
    {
        \includegraphics[width=1.91in ,angle=270]{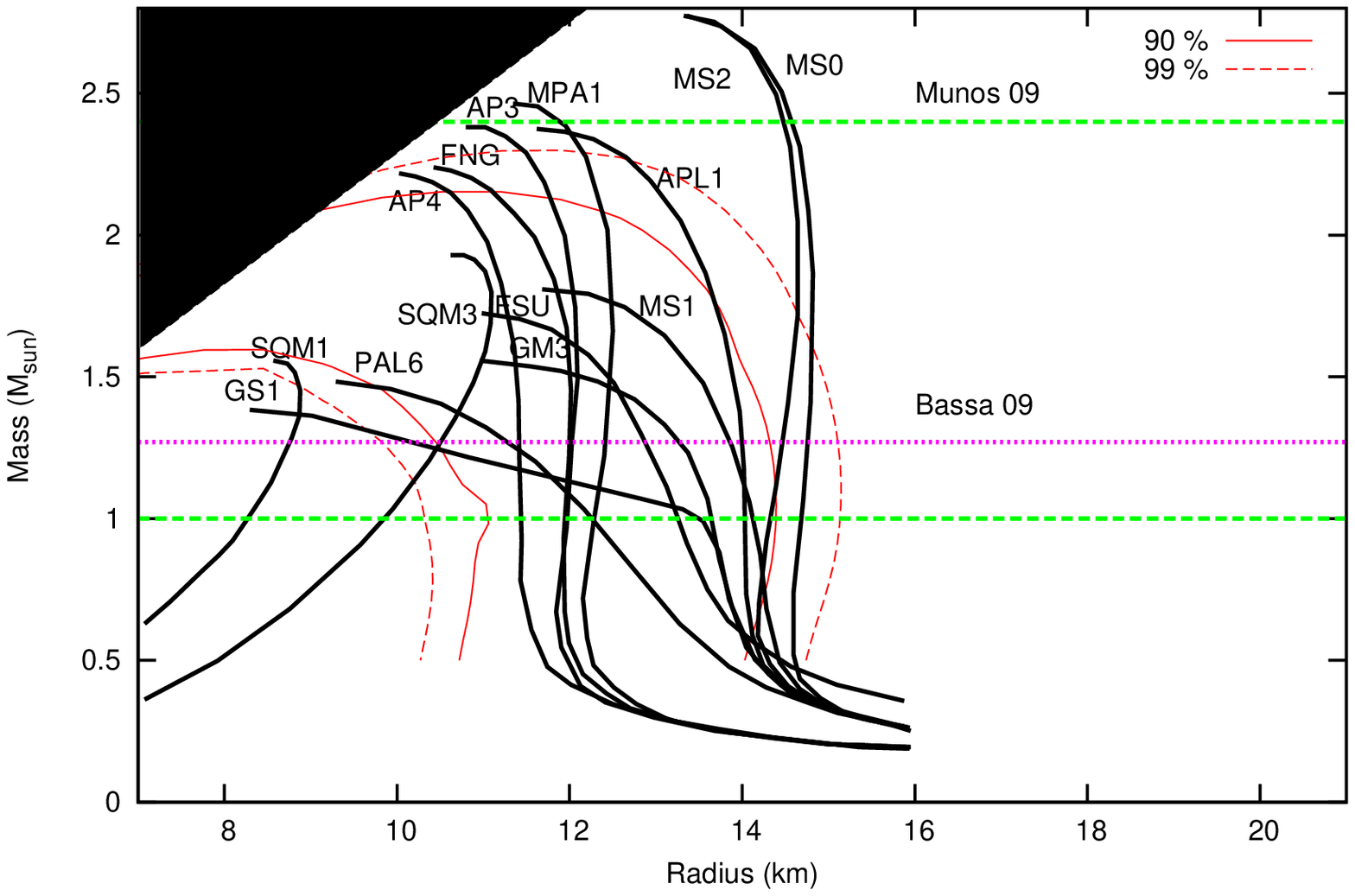}
        \label{fig:second_sub}
    }
    \subfigure[6.0 kpc]
    {
        \includegraphics[width=1.91in,angle=270]{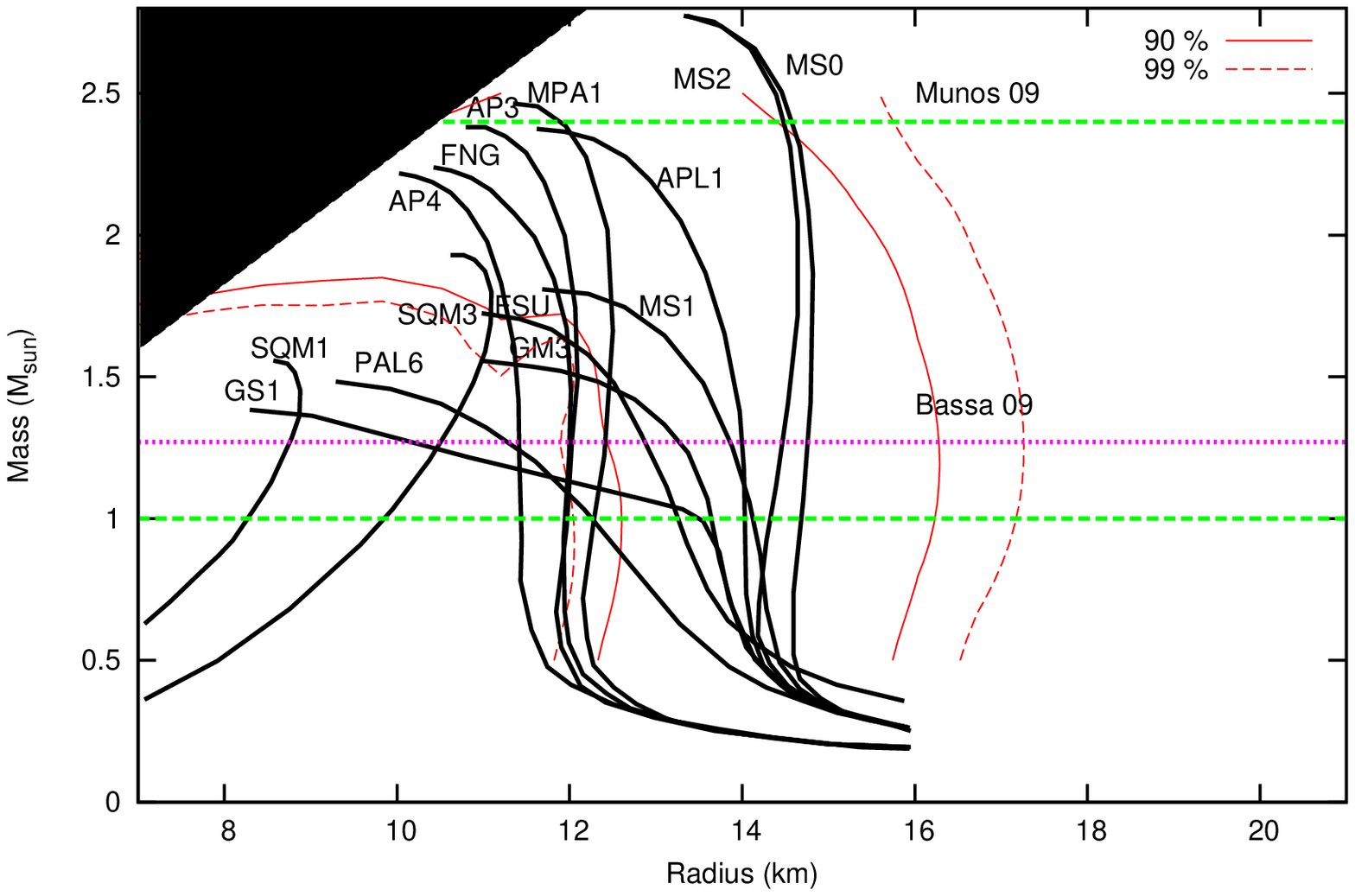}
        \label{fig:third_sub}
    }
    \subfigure[7.0 kpc]
    {
        \includegraphics[width=1.91in,angle=270]{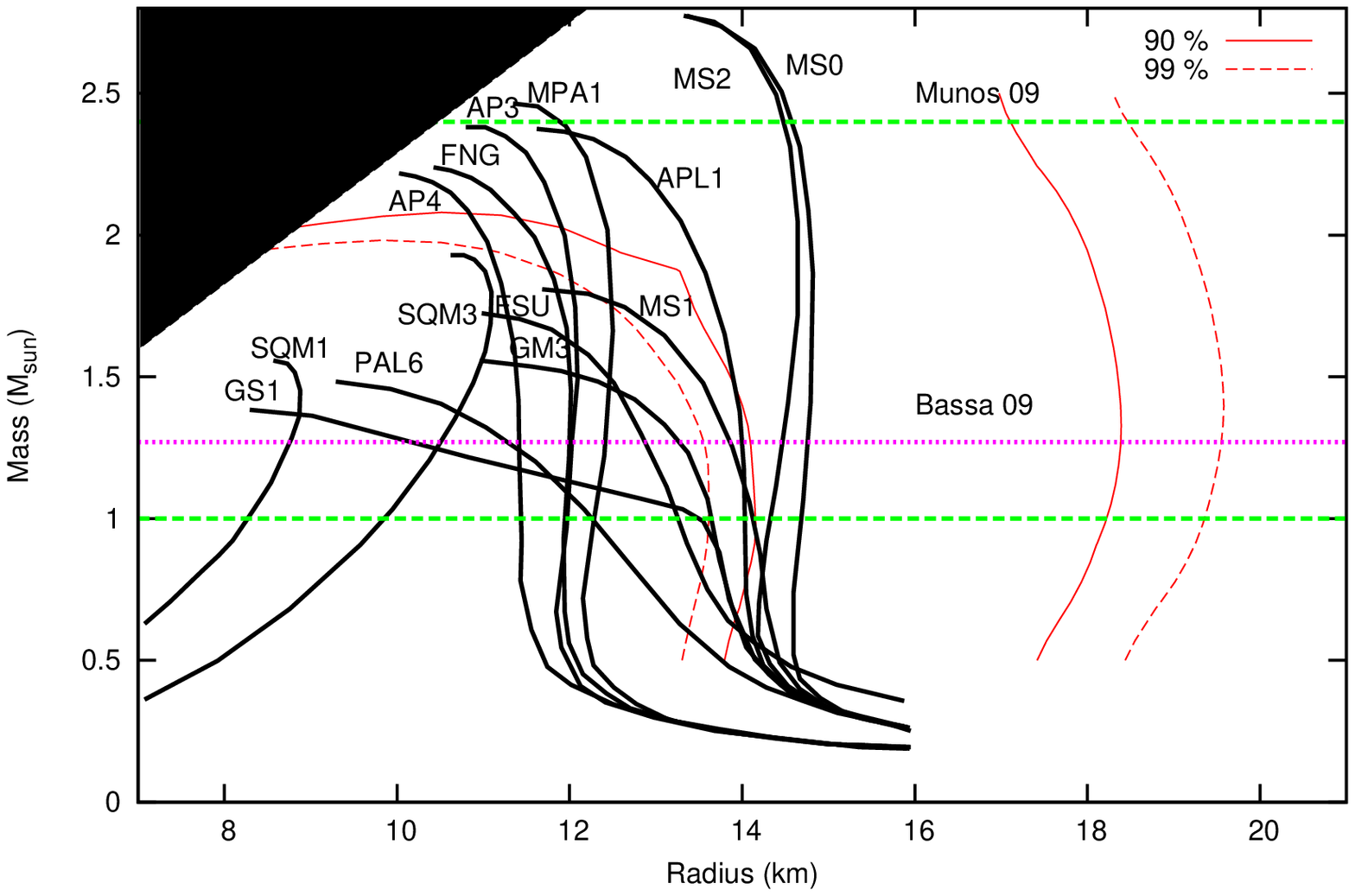}
        \label{fig:second_sub}
    }
    \subfigure[8.0 kpc]
    {
        \includegraphics[width=1.91in,angle=270]{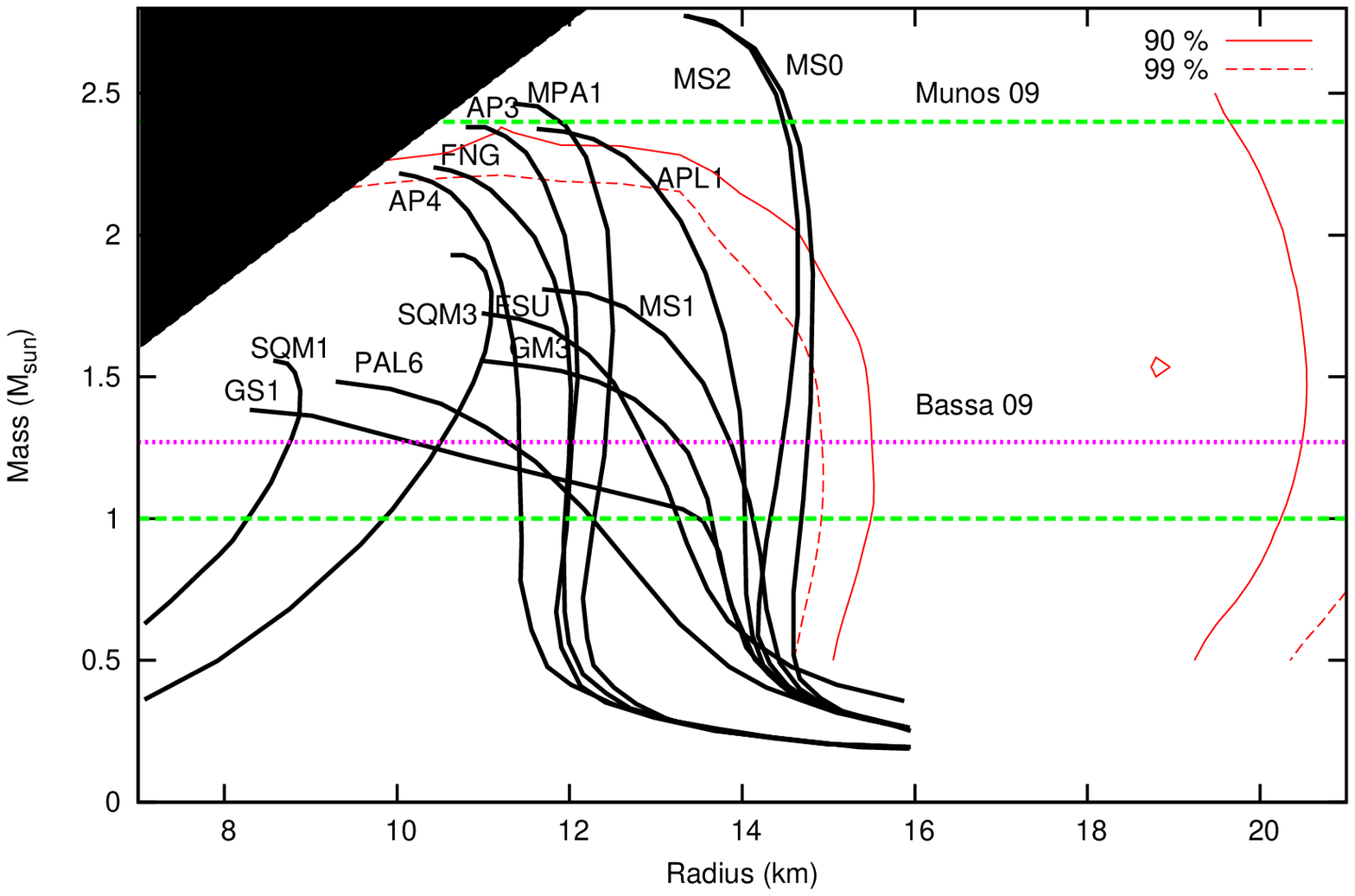}
        \label{fig:third_sub}
    }
    \subfigure[9.0 kpc]
    {
        \includegraphics[width=1.91in,angle=270]{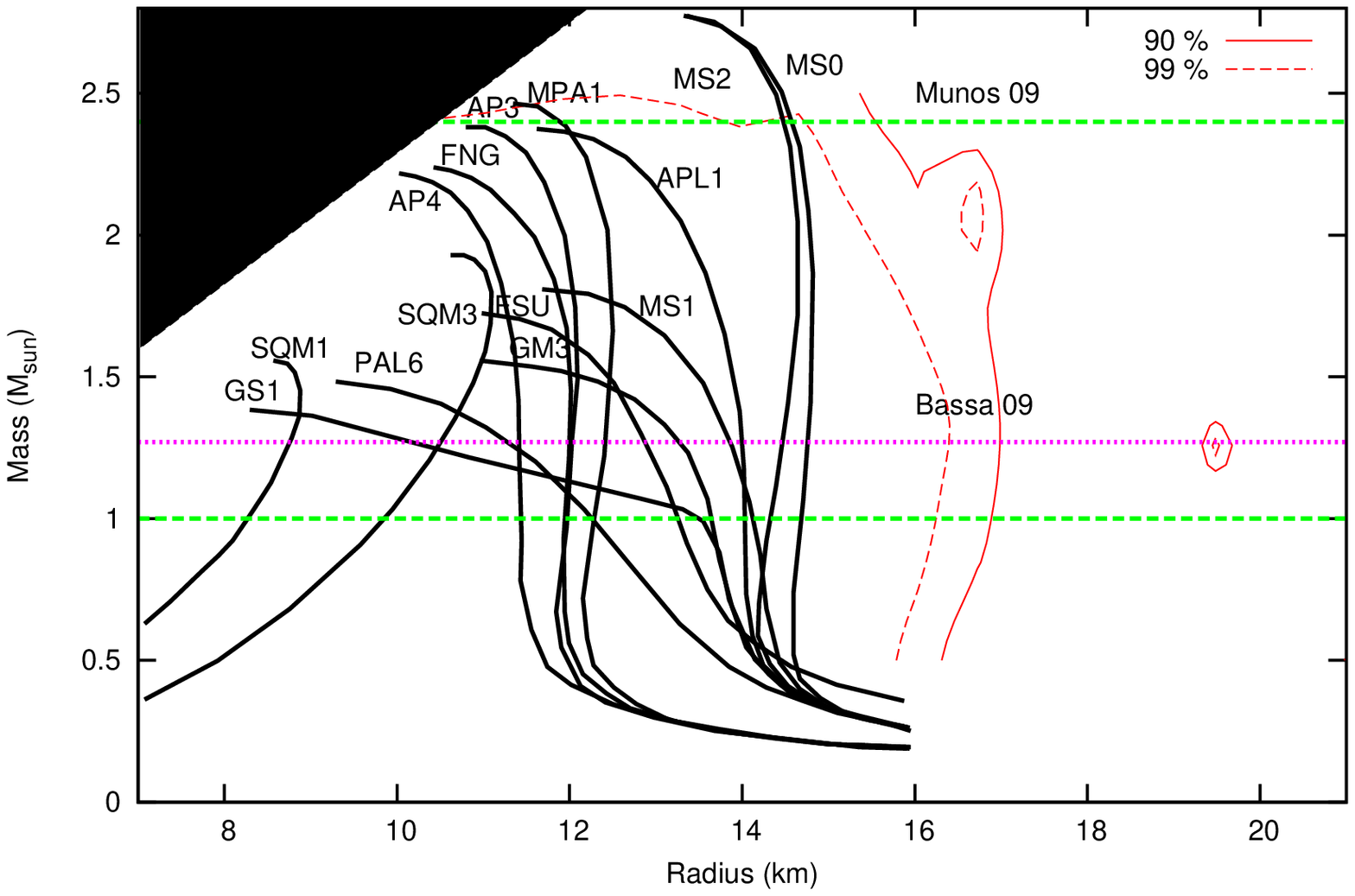}
        \label{fig:second_sub}
    }
    \subfigure[10.0 kpc]
    {
        \includegraphics[width=1.91in,angle=270]{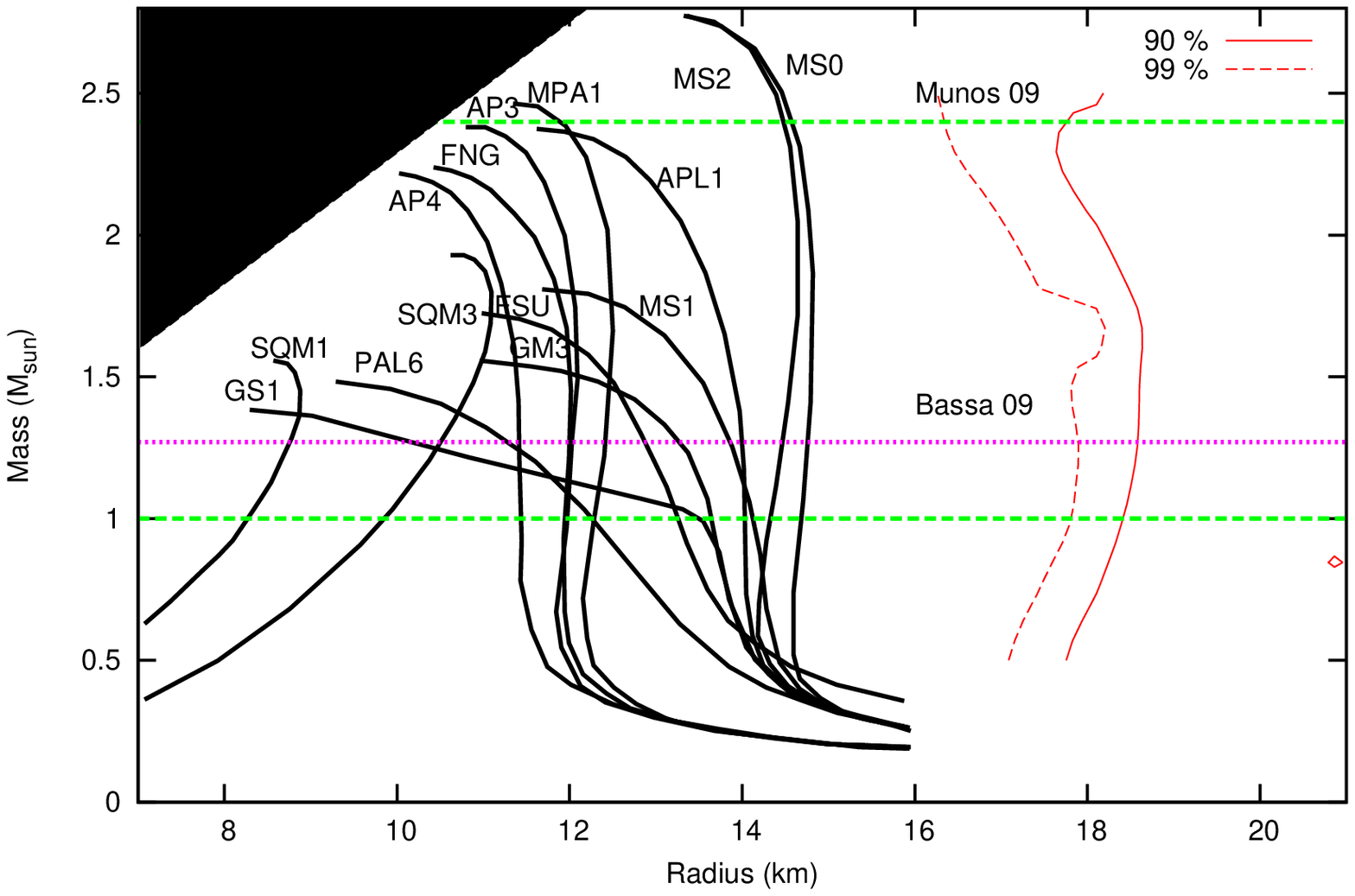}
        \label{fig:third_sub}
    }
\caption{Contour plots showing the results of modeling the neutron-star in \EXO with the
{\sc xspec} model NSATMOS and power-law. The power-law index is fixed to 1.  The
plots show two confidence levels in the mass-radius diagram obtained from our fit;
the contour lines (red) are for the confidence levels of 90\% and 99\%, respectively.
The pink line ``A'' is for the lower limit of $M_{\rm NS}$ given by Bassa et al.(2009), and 
the green line ``B'' is for the upper limit given by Mu\~noz et al. (2009). }
\label{fig:contour_sum}
\end{center}
\end{figure*}

We used the {\sc steppar} command in {\sc xspec} to vary the mass, radius and distance 
parameters simultaneously, allowing other parameters to be free to find the best fit at each
step.  For the mass we go from 0.5 to 2.5 $M_{\odot}$ with steps of 0.1 $M_{\odot}$, and for the distance we go from 5 to 10 kpc with steps of 0.25 kpc.
The minimum and maximum radius allowed with these
models are 5.0 km and 25.0 km, respectively. 
In Fig \ref{fig:contour_sum} we show the contour plots obtained from the STEPPAR procedure for the NSATMOS model. Each plot is for a different distance, ranging from 5 to 10 kpc. The contour lines (red) are for the confidence levels of 90\% (solid) and 99\% (dashed). Further, in Fig \ref{fig:contour_sum} we give different neutron-star EOS (black) taken from \cite{Lattimer07}.
We did the same analysis for the NSAGRAV model as well.
Both two models give consistent result, in accordance with the findings of \cite{Webb07}.

Using the optical data from the Very Large Telescope (VLT), moderate-resolution spectroscopy
of the optical counterpart and Doppler tomography, \cite{Munoz09} provided the first
dynamical constraints on the stellar mass of LMXB \EXO.  The mass range of the
neutron star that they derived is $1 \Msun \le {\rm M_{\rm NS}} \le 2.4 \Msun$.
Subsequently, \cite{Bassa09} analyzed  optical spectra of \EXO when the source
was in the  quiescent state, and they gave a lower limit to the neutron-star mass of 
${\rm M_{\rm NS}} \ge $ 1.27 $\Msun$. As upper limit we used the value  reported by
\cite{Munoz09}, but since at the time of their observation the source was still in outburst,
we used the lower limit  reported in \cite{Bassa09}. In Figure \ref{fig:contour_sum} we also give the lower (pink/dotted) and upper (green/dashed) limits to the neutron-star mass.

In order to test the EOS  and identify the upper limit to the source distance, 
we assume three different EOS models: normal  nucleonic matter  (AP3), 
boson condensates matter (MS1) and strange quark matter  (SQM1).
By varying the source distance from 5 to 10 kpc, the contour lines for the fitted model
 move on the NS mass-radius diagram. We can estimate the probability 
of the distance for each EOS when the contour lines pass through the EOS curves. 
Note that as the distance increases (see Figure \ref{fig:contour_sum}), the satisfied area of
the model moves from bottom left to top right in the plot. The  results using NSAGRAV 
 are similar to those shown in Figure \ref{fig:contour_sum}. 
For a certain distance we found that not all the EOSs are consistent with the two neutron-star atmosphere models that we used.

If the neutron star in \EXO follows the EOS model `AP3',
the probability that the source has a distance of 10.0 kpc is $1 \times
10^{-4}$ and $1 \times 10^{-6}$ for NSAGRAV and NSATMOS, respectively. 
If we want to get a probability for the distance larger than $1 \times 10^{-2}$ 
(99\% confidence), the distance for NSAGRAV and NSATMOS should be smaller than
8.9 kpc and 8.5 kpc, respectively. The distance at 90\% confidence for NSAGRAV 
and NSATMOS is less than 8.3 kpc and 8.2 kpc, respectively. Both models are consistent with the distance of 7.1 kpc given by type-I X-ray bursts \citep{Galloway08b}.

For the EOS model 'MS1', the probabilities that \EXO is at a distance
of 10 kpc is $10^{-5}$ and $10^{-6}$ for NSAGRAV and NSATMOS, respectively. 
For both models, respectively,   the distance at 99\% confidence level is less than 7.3 kpc 
and 7.1 kpc, and the distance at 90\% confidence level is less than 6.9 kpc and 6.8 kpc. 
Both neutron-star atmosphere models with the 'MS1' model have an upper limit for the  distance 
smaller than 7.1 kpc. 

For a EOS model `SQM1', the distance at 99\% confidence level is less
than 5.2 kpc, and the distance at 90\% confidence level is less than 5.0 kpc for both atmosphere models. The upper limits on the distance to \EXO for different 
EOS  are shown in Table \ref{tab:upper_limit}. The 'SQM1' model is rejected at a 99\% confidence level for 
this neutron star, unless the source is closer than 5.2 kpc. 

\begin{table}
\caption{Upper limits on the distance to \EXO for different EOS models. }

\begin{tabular}{ccccccccccc}

\hline \hline
EOS         & AP3   & AP3   & MS1   & MS1   & SQM1 & SQM1  \\
\hline 
confidence  & 90\%  & 99\%  & 90\%  & 99\%  & 90\% & 99\% \\
\hline
NSAGRAV     & $<$ 8.3 & $<$ 8.9  & $<$ 6.9  & $<$ 7.3 & $<$ 5.0 & $<$ 5.2 \\
\hline 
NSATMOS     & $<$ 8.2 & $<$ 8.5  & $<$ 6.8  & $<$ 7.1 & $<$ 5.0 & $<$ 5.2 \\

\hline 
\label{tab:upper_limit}
\end{tabular}
\begin{tablenotes}
\item[]Note. --The 90\% and 99\% confidence levels upper limit for the two NS atmosphere models  
NSAGRAV and NSATMOS for the EOS models: `AP3', `MS1' and 'SQM1'. The distance is in kpc. 
\end{tablenotes}
\end{table}

\section[]{DISCUSSION}
\label{discussion}

We analyzed an XMM-Newton observation of the neutron star \EXO in the quiescent
state. The unabsorbed X-ray flux in the 0.5$-$10.0 keV energy band was $\sim 1.1\times$
10$^{-12}$ ergs cm$^{-2}$ s$^{-1}$. We found that the non-thermal (power-law) component 
only contributes  $\sim 10\pm 2 \%$ of the 0.5$-$10 keV X-ray flux, which is lower than what
\cite{Degenaar09} found from {\em Chandra} data ($F_{pow}$ was $\sim$ 16$-$17$\%$ of  
the 0.5$-$10 keV X-ray flux from the fit with $\Gamma=1$) about a month earlier than our observation. The total 
unabsorbed flux (0.5$-$10.0 keV) decreased from $1.3\times 10^{-12}$ ergs cm$^{-2}$ s$^{-1}$ 
in the {\em Chandra} observation to $1.1\times 10^{-12}$ ergs cm$^{-2}$ s$^{-1}$ in our observation,
whereas $N_{\rm H}$ changed from $\sim 1.2$ $\times 10^{21}$ $\rm cm^{-2}$ to $\sim 0.6$ $\times 10^{21}$ $\rm cm^{-2}$. The effective temperature, however, did not show large variations in one month time. According to the above 
comparisons, the reduction of the total flux is due to a lower contribution
of the power-law component. 

Because the X-ray spectrum in the quiescent state is dominated by thermal emission
originating from the NS surface, our data allow us to  constrain the  mass and 
radius of the neutron star. From the two different NS atmosphere models (NSAGRAV and NSATMOS) that we used to fit the X-ray spectrum, we found that both models show similar results and set good constraints on the neutron-star radius. Even taking into account  the $M_{\rm NS}$ lower limit  \cite[from][]{Bassa09},  upper limit  \cite[from][]{Munoz09} and our best fit $\Delta \chi^2$ contour, we still have a large area 
on the mass-radius diagram, and many EOSs are still possible (see Figure \ref{fig:contour_sum}). In order to constrain the allowed space of mass and radius at a specified distance, we choose three typical neutron-star EOS, `AP3', `MS1' and 'SQM1'. We found that the smaller the distance to 
the NS the more EOSs are consistent with the data.

For any specific EOS, as the upper limit of the distance we took the value of the distance where the 99\% confidence contour just intersects the curve of that EOS.
We found that the upper limits on the distance as derived from the NSAGRAV model are
slightly higher than those for the NSATMOS model. The EOS model `MS1' 
can be just satisfied at a  distance of 7.1 kpc. If we assume that the neutron 
star in \EXO is a normal neutron star, following the EOS `AP3', the source should be
closer than 8.9 kpc for the NSAGRAV model,  or 8.5 kpc for the NSATMOS model. 
Both the  'MS1' and   'AP3' EOS  are fully consistent with  the measured distance 
of 7.1 kpc \citep{Galloway08b, Wolff05}. 
For larger distances more EOS are ruled out.  The EOS 'SQM1' is 
rejected by the atmosphere model fits for a  distance of 7.1 kpc measured 
from the X-ray bursts \cite{Galloway08b}. We note, however, the neutron-star atmosphere models may not
appropriate for 'bare' quark matter stars, but only for those normal quark star where a crust is present.

\section*{Acknowledgments} 

This work is based on the observations obtained from XMM-Newton. PGJ acknowledges support from a
VIDI grant from the Netherlands Organisation for Scientific Research.


\label{lastpage}

\end{document}